# RECONSTRUCTION OF HIGH-ENERGY PART OF THE GAMMA-RAY SPECTRUM IN THERMAL NEUTRON CAPTURE BY $^{113}$Cd

V. A. Plujko[1,2,*], O. M. Gorbachenko[1], K. M. Solodovnyk[1], V. M. Petrenko[1]

[1] *Taras Shevchenko National University of Kyiv, Kyiv, Ukraine*
[2] *Institute for Nuclear Research, National Academy of Sciences of Ukraine, Kyiv, Ukraine*

*Corresponding author: plujko@gmail.com

The average gamma-ray spectrum $^{114}Cd$ after thermal neutron capture in $^{113}Cd$ was evaluated in units of mb/MeV. Two approaches are considered for estimation of the average gamma-ray spectrum with normalization of the experimental data: mean spectra for all gamma-energies were found by averaging frequency polygon for experimental data histogram, and mean spectra were estimated as the combination of theoretical values at low gamma-ray energies and averaging experimental data in high-energy range. The experi mental spectra were evaluated from the gamma-intensities presented by Mheemeed et al. [A. Mheemeed et al., Nucl. Phys. A 412 (1984) 113] and Belgya et al [T. Belgya et al., EPJ Web of Conf. 146 (2017) 05009]. They were normalized to the average theoretical spectrum which was calculated using EMPIRE and TALYS codes. The procedure of normalization of the high-energy part of the spectrum was described. Estimated $\gamma$-spectra for $^{113}Cd(n,\{x\gamma\})$ reaction induced by thermal neutrons were presented.

## 1. Introduction

The cross-sections of gamma-ray production induced by neutron interactions are required for the development of the reactor technologies, such as radiation damage of reactor construction elements undergoing neutrons irradiation, shielding calculations of the reactors [1, 2], *etc*. Investigations of nuclear reactions with neutrons of low energies are important for these tasks because cross-sections of the interaction with neutrons are rather high. The cadmium element is widely used in the reactor industry as the neutron absorber. It can be noted [3] that gamma-spectrum from thermal neutrons is also needed for accurate estimation of gamma spectrum for the fast neutrons in indoor experiments due to the possibility of re-scattering fast neutrons on the experimental facilities and surroundings down to the thermal energies.

For thermal neutron capture in $^{113}Cd$, the transition intensities, $I(E_\gamma)$, of the gamma-rays were measured in Refs. [4,5]. They were presented in units of the number of gamma-quanta per 10000 captured neutrons. In the applications, as a rule, the normalized values of the gamma-ray spectrum $d\sigma_\gamma(E_n, E_\gamma)/dE_\gamma$ are needed in terms of mb/MeV. Here we reconstruct, in an average, the gamma-spectra induced by thermal neutrons [4,5] in $^{113}Cd$ in such units. Normalization is done by the use of theoretical calculations of the gamma-ray spectrum of $^{113}Cd(n,\{x\gamma\})$ reaction performed by the use of EMPIRE 3.2.2 and TALYS 1.6 codes [6, 7]. The procedure of normalization is described in Sect 2.

## 2. Evaluation of gamma-ray spectrum

Fig. 1 presents the experimental intensities $I(E_\gamma)$ of the gamma rays following thermal neutron capture in $^{113}Cd$ from refs. [4,5] in units of the number of gamma-quanta per 10000 captured neutrons as a function of the gamma-ray energy $E_\gamma$. The experimental data from Ref. [4] were taken from Table A1 which summarizes the observed properties of transitions in $^{114}Cd$ following thermal neutron capture. These data were not averaged in comparison with data from Ref. [5] which were binned to 1 keV/bin. Therefore, the data from Ref. [4] presented in Fig. 1 lie higher (specifically in the high-energy range) than the data from Ref. [5] but their average values have similar behavior.

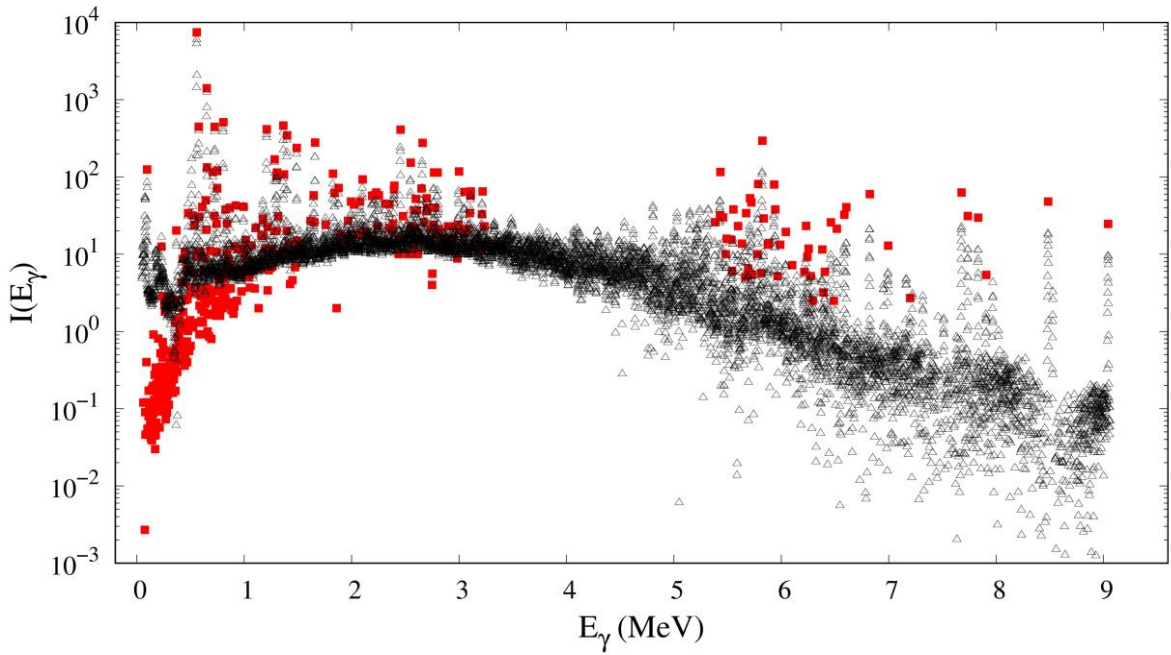

Fig. 1. The experimental intensities of gamma-ray transitions in $^{114}Cd$ following thermal neutron capture. Experimental data are taken from [4] (squares) and [5] (triangles).

The gamma-ray spectra in thermal neutron capture by $^{113}Cd$ can not be directly calculated by TALYS and EMPIRE codes. Minimal neutron energy $E_{n,\min}$ at which such gamma-ray spectra can be calculated within the TALYS code is $E_{n,\min} = 1 eV$ and $E_{n,\min} = 0.55$ eV for EMPIRE. Therefore we evaluate the spectra for thermal neutron with energy $E_n = E_{th} = 0.0253 eV$ using several additional assumptions. Initially the calculations within the codes were performed at the same neutron energy $E_n^* = 1 eV$, and then the theoretical gamma-ray spectra for neutron thermal energy $E_{th} = 25.3 meV$ were recalculated by the following scaling approximation

$$\frac{d\sigma_\gamma^{(\alpha)}(E_n = E_{th} = 0.0253 eV, E_\gamma)}{dE_\gamma} = K^{(\alpha)} \cdot \frac{d\sigma_\gamma^{(\alpha)}(E_n = E_n^* = 1.0 eV, E_\gamma)}{dE_\gamma}, \quad (1)$$

where index $\alpha$ denotes gamma-spectra within EMPIRE and TALYS codes ($\alpha = E, T$) in terms of mb/MeV and $K^{(\alpha)}$ is a scaling factor for calculation within code $\alpha$.

The scaling approximation (1) is based on the following assumptions: I) neutron capture proceeds through the compound nucleus mechanism and mean gamma-ray spectrum can be presented as the product of the neutron capture cross-section $\sigma_\gamma^{(\alpha)}(E_n)$ into probability $G_\gamma^{(\alpha)}(E_\gamma)$ of gamma-decay in a unit of gamma-ray energy; II) the probability $G_\gamma(E_\gamma)$ of gamma-decay is almost independent of neutron energy $E_n$ in the range from neutron thermal energy untill 1 eV. The scaling factor $K^{(\alpha)}$ under assumptions I, II is given by the following formula

$$K^{(\alpha)} = \frac{\sigma_\gamma(E_n = E_{th} = 0.0253 eV)}{\sigma_\gamma^{(\alpha)}(E_n = E_n^* = 1.0 eV)}.$$

For cross-section of thermal neutron capture we adopt the value $\sigma_{th} \equiv \sigma_\gamma(E_n = E_{th} = 0.0253 eV) =$ 19964.1 b from library ENDF/B-VIII.0. which is in close agreement with these ones from all other libraries (TENDL-2019 (WEB-site of NDS-IAEA), BROND-3.1 etc). The values of $\sigma_\gamma^{(T)}(E_n = E_n^*) = 247.9$ b and $\sigma_\gamma^{(E)}(E_n = E_n^*) = 235.2$ b at neutron energy 1 eV were calculated by TALYS and EMPIRE codes with default expressions for dipole electric photon strength function (PSF) and for nuclear level density (NLD) model [6-8], namely, EMPIRE code: MLO1 ((Modified Lorentzian 1, GSTRFN=1) for PSF and EGSM (Empire-specific level density, LEVDEN=0) for NLD; TALYS code: EGLO(Enhanced generalized Lorentzian, strength=1) for PSF and GC(Gilbert-Cameron model, ldmodel=1) for NLD.

Figs. 2 and 3 show recalculated gamma-ray spectra for $(n,\{x,\gamma\})$ reactions in $^{113}Cd$ for the neutrons with thermal energy. The calculations with different dipole PSF are presented on the left panels and the results using different nuclear level density NLD models are demonstrated on right panels.

The following PSF models were used [8] for the calculations within the EMPIRE code: MLO1 (Modified Lorentzian 1, GSTRFN=1), EGLO (Enhanced generalized Lorentzian, GSTRFN=0), SLO (Standard Lorentzian, GSTRFN=6), GFL (Generalized fluid liquid approach, GSTRFN=5). Such NLD models were used: EGSM (Empire-specific level density, LEVDEN=0), GSM (Generalized Superfluid Model, LEVDEN=1), HFBM (microscopic combinatorial Hartree–Fock–Bogolyubov model, LEVDEN=3).

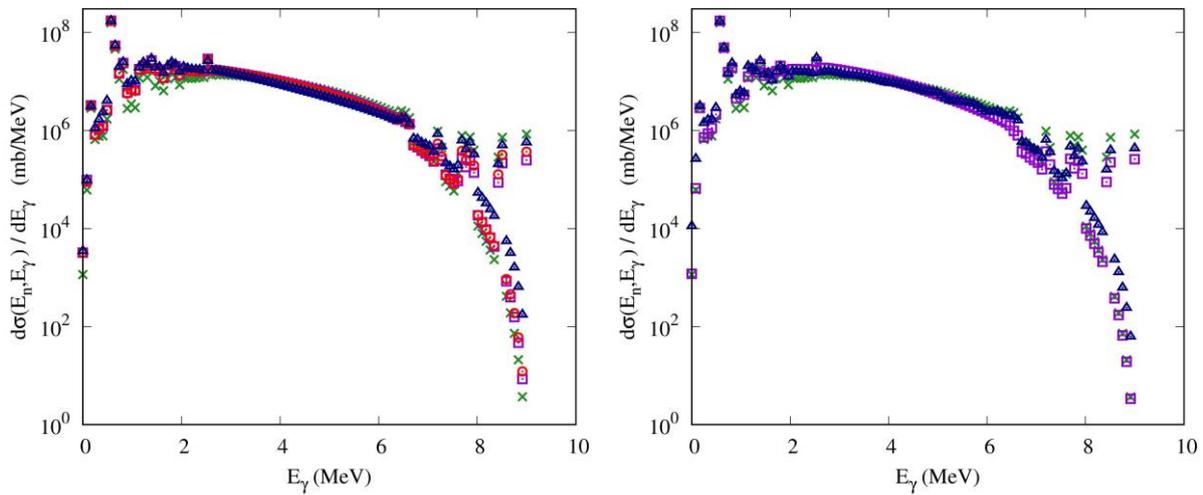

Fig. 2. The theoretical calculations of differential gamma-ray spectra induced by the thermal neutrons on $^{113}Cd$ calculated with the use of EMPIRE code. The left panel shows results for different modes of photon strength functions with the EGSM model for NLD: MLO (crosses), SLO (open squares), GFL (open circles), EGLO (open triangles). The right panel shows results for different nuclear level densities models with the MLO for PSF: EGSM (crosses), GSM (open squares), HFBM (open triangles); $K^{(E)} = 84.9$.

The calculations within the TALYS code were performed using the SLO(strength=2) and EGLO(strength=1) models for the PSF and GSM(ldmodel=3), FG (Fermi gas model(ldmodel=2)), GC (Gilbert-Cameron model(ldmodel=1)) for the NLD [8].

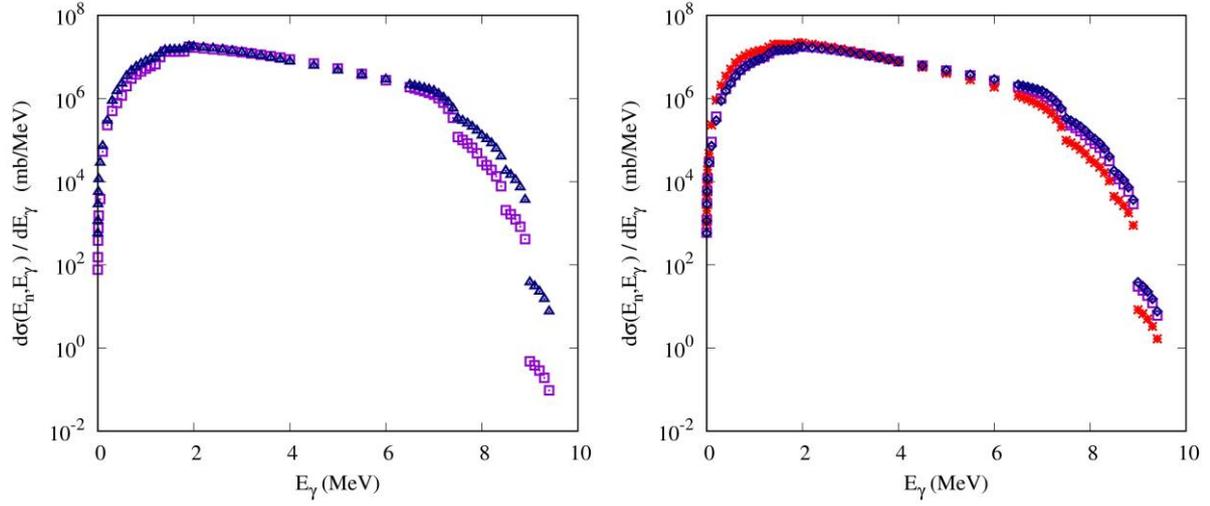

Fig. 3. The theoretical calculations of gamma-ray spectra induced by the thermal neutrons on $^{113}$Cd calculated with the use of the TALYS code. The left panel shows results for different models of photon strength functions: SLO (open squares), EGLO (open triangles). The right panel shows calculations for different nuclear level densities models: GSM (open squares), FG (stars), GC (diamond); $K^{(T)} = 80.5$.

These figures demonstrate rather slight dependences of the theoretical gamma-ray spectra on the models of the PSF and NLD in the gamma energy range from ~ 2.0 MeV until ~ 6.5 MeV. It can be seen by comparison figs.2 and 3, that, in the low gamma energy range till ~ 2.0 MeV, the shape of the theoretical calculations by the EMPIRE code agrees better with the experimental data. For the high energy range above > 6.5 MeV, theoretical spectra are strongly dependent on the form of the PSF and NLD. There is the possibility that results within the EMPIRE code can describe the high-energy part of the spectrum after modifications of expressions for PSF and NLD.

In order to evaluate the most reliable gamma-ray spectra induced by thermal neutrons, we use experimental data on gamma-transition intensities normalized to average theoretical spectra, which are calculated by EMPIRE and TALYS codes and scaling approximation. In the experiments [4,5] with gamma-ray emission, the values of transition intensities $I(E_\gamma = \Delta E_m) = I(E_m)$ between discrete levels with differences $\Delta E_m = E_{m+1} - E_m$ of gamma-ray energies are measured, where $E_m$ and $E_{m+1}$ are the energies of final and initial states. The intensities $I_m$ are presented as variation series, i.e., as numbers of intensities in order of increasing values of the energy interval $\Delta E_1 < \Delta E_2 < \Delta E_3 ...$.

Mean gamma-ray spectrum $d\bar\sigma(E_\gamma)/dE_\gamma$,

$$\frac{d\bar{\sigma}(E_\gamma)}{dE_\gamma} \equiv \int w(E_\gamma, E_\gamma')\frac{d\sigma(E_\gamma')}{dE_\gamma'}dE_\gamma' = \frac{1}{h}\int_{E_\gamma-h/2}^{E_\gamma+h/2}\frac{d\sigma(E_\gamma')}{dE_\gamma'}dE_\gamma' , \qquad (2)$$

averaged by the weight function $w(E_\gamma, E_\gamma')$ of the rectangular shape of the width $h$, is proportional to the histogram $H_h(E_\gamma)$ of intensities $I$ of gamma transitions per bin width $h$ for the interval from energy $E_\gamma - h/2$ till $E_\gamma + h/2$ with mean gamma-ray energy $E_\gamma$

$$\frac{d\bar{\sigma}(E_\gamma)}{dE_\gamma} = F \cdot H_h(E_\gamma) . \quad H_h(E_\gamma) = \frac{1}{h}\sum_{j=1}^{N_\gamma} I \ (E_\gamma - h/2 \leq E_{\gamma,j} \leq E_\gamma + h/2) , \qquad (3)$$

where $F$ is the normalization factor and $N_\gamma$ is the number of data points in interval $h$.

The factor $F$ is found by comparison of total theoretical gamma-ray spectra, $S_{k,the}^{(\alpha)}$, in some energy interval $h_k$ calculated within EMPIRE and TALYS codes ($\alpha = E$ and $T$):

$$S_{k,the}^{(\alpha)} = \int_{\Delta E_k}^{\Delta E_k + h_k} \frac{d\sigma_\gamma^{(\alpha)}(E_{th}, E_\gamma)}{dE_\gamma} dE_\gamma , \qquad (4)$$

with the experimental analogue of total gamma spectra $S_{k,\exp}^{(\alpha,\varepsilon)}$,

$$S_{k,\exp}^{(\alpha,\varepsilon)} = F_k^{(\alpha,\varepsilon)} \cdot i_k^{(\varepsilon)} , \quad i_k^{(\varepsilon)} = \sum_{j=1}^{N_k^{(\varepsilon)}} I^{(\varepsilon)}(E_{\gamma,j} \in \Delta E_k^{(\varepsilon)} \div \Delta E_k^{(\varepsilon)} + h_k) , \qquad (5)$$

where index $\varepsilon = M, B$ denotes experimental data from the refs. [4] or [5] respectively; $E_{\gamma,j}$ ($j = 1 - : - N_k^{(\varepsilon)}$) are the experimental gamma-ray energies with the number of data points $N_k^{(\varepsilon)}$ and $i_k^{(\varepsilon)}$ is a sum of the gamma-ray intensities in the energy range $\Delta E_k^{(\varepsilon)} \div \Delta E_{k+1}^{(\varepsilon)}$; $F_k^{(\alpha,\varepsilon)}$ are the normalization factors. Additional indexes $\alpha, \varepsilon$ are introduced for quantities in equations (4) and (5) for using different data and codes.

The normalization factors are found using equality of the relationships (4) and (5), i.e. by condition $S_{k,the}^{(\alpha)} = S_{k,\exp}^{(\alpha)}$:

$$F_k^{(\alpha,\varepsilon)} = S_{k,\exp}^{(\alpha)} / i_k^{(\varepsilon)} . \qquad (6)$$

For the calculation of the normalization factors $F_k^{(\alpha,\varepsilon)}$ by eq. (6), two ranges of gamma-ray energies $\Delta E_A$ and $\Delta E_B$ are taken, namely, $\Delta E_A = 1879.10 \div 3238.6$ keV and $\Delta E_B = 5384.9 \div 6605.4$ keV with intervals $h_A = 1359.5$ keV and $h_B = 1220.5$ keV respectively. These energy ranges include many gamma transitions that vary rather slightly. Theoretical gamma-ray spectra in these energy ranges depend also slightly on the models of NLD and PSF. Because of this, we use these intervals for the calculation of normalization factors $F_k^{(\alpha,\varepsilon)}$ where it is expected that their values will only slightly be dependent on the models. The calculations with default expressions for PSF and NLD are used (above-listed). The intervals $\Delta E_A$ and $\Delta E_B$ are separated into two subintervals denoted as $\Delta E_{A1}$, $\Delta E_{A2}$ and $\Delta E_{B1}$, $\Delta E_{B2}$. The normalization constants were determined by equation (6) for each interval. The arithmetical means of the constants are taken for normalization factors $F^{(\alpha,\varepsilon)} = (F^{(\alpha,\varepsilon)}_{\Delta E_{A1}} + F^{(\alpha,\varepsilon)}_{\Delta E_{A2}} + F^{(\alpha,\varepsilon)}_{\Delta E_{B1}} + F^{(\alpha,\varepsilon)}_{\Delta E_{B2}})/4$ with the results $F^{(E,M)} = 99.36$, $F^{(E,B)} = 23.80$, $F^{(T,M)} = 60.34$, $F^{(T,B)} = 13.99$.

We consider two variants for estimation of the average gamma-ray spectrum with normalization of the experimental data; I) mean spectrum for all gamma-energies is found by averaging frequency polygon $P_h(E_\gamma)$ for experimental data histogram $H_h(E_\gamma)$, and II) mean spectrum is estimated as the combination of theoretical values at low gamma-ray energies and averaged experimental data in high-energy range with $E_\gamma > 6.5$ MeV.

For calculation of mean gamma-spectrum $d\bar{\sigma}_P(E_\gamma)/dE_\gamma$ from frequency polygon, the weight function $w(E_\gamma, E_\gamma')$ of averaging is taken in the rectangular shape with the width $h$, that is, similarly to equation (2), the following relationship is used

$$\frac{d\bar{\sigma}_P(E_\gamma)}{dE_\gamma} \equiv F \cdot \int w(E_\gamma, E_\gamma') P_h(E_\gamma') dE_\gamma' = \frac{F}{h} \int_{E_\gamma - h/2}^{E_\gamma + h/2} P_h(E_\gamma') dE_\gamma' \,. \qquad (7)$$

Fig. 4 demonstrates the dependences of normalized frequency polygons $Pn^{(\alpha,\varepsilon)}(E_\gamma) = F^{(\alpha,\varepsilon)} \cdot P_h^{(\varepsilon)}(E_\gamma)$, mean gamma-spectra $d\bar{\sigma}_P^{(\alpha,\varepsilon)}/dE_\gamma$, and the theoretical spectra $d\sigma_\gamma^{(\alpha)}/dE_\gamma$ on gamma-ray energy using experimental data by Mheemeed et al [4] and Belgya et al [5] ($\varepsilon = M, B$), calculations within EMPIRE and TALYS codes ($\alpha = E, T$) and scaling approximation. The left panel shows evaluations based on the experimental data from [4] and the right panel - data from [5]. In Figs.4,5 the curves were calculated with expressions for NLD and PSF by default, namely, EMPIRE code: EGSM (LEVDEN=0) for NLD and MLO1(GSTRFN=1) for PSF; TALYS code: GC(ldmodel=1) for NLD and EGLO(strength=1) for PSF.

In equations (3), (7), the width $h$ was taken from the beginning of every interval in two variants: 1) $h = 0.5$ MeV and 2) $h = 0.8$ MeV. The lengths of the last widths on the right edges were equal to the total length of intervals minus the product of the values $h$ and the maximal number of integer values $h$ in the intervals.

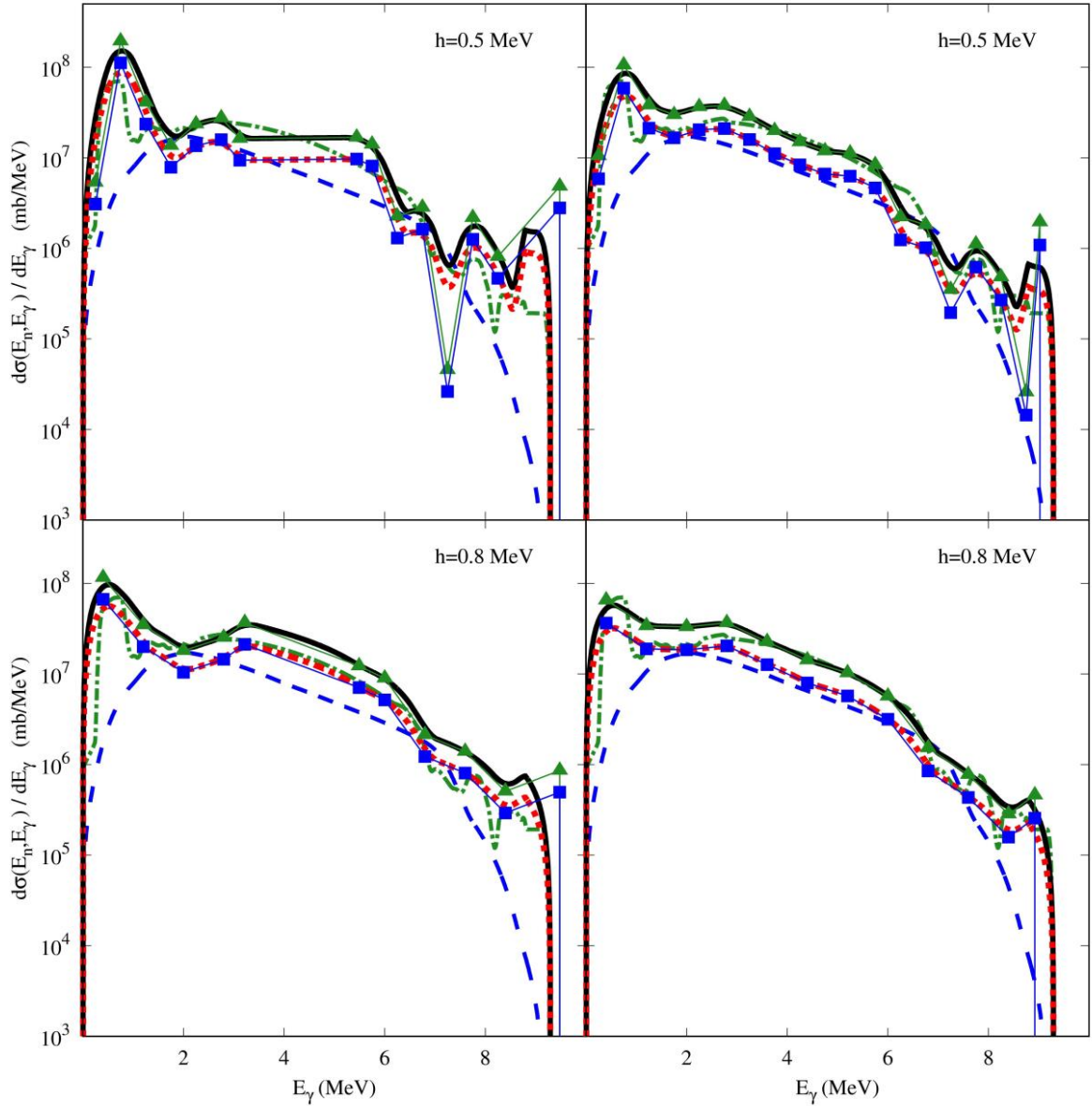

Fig. 4. Dependences of normalized frequency polygons $Pn^{(\alpha,\varepsilon)}(E_\gamma) = F^{(\alpha,\varepsilon)} \cdot P_h^{(\varepsilon)}(E_\gamma)$, mean gamma-spectra $d\bar{\sigma}_P^{(\alpha,\varepsilon)}/dE_\gamma$, and the theoretical spectra $d\sigma_\gamma^{(\alpha)}/dE_\gamma$ on gamma-ray energy using different experimental data and calculations within EMPIRE and TALYS codes ($\alpha = E, T$). The left panel shows evaluations based on the experimental data by Mheemeed et al [4] ($\varepsilon = M$), the right panel - evaluations using data Belgya et al [5] ($\varepsilon = B$). The denotations: dash-dot line presents $d\sigma_\gamma^{(E)}/dE_\gamma$, dashed line - $d\sigma_\gamma^{(T)}/dE_\gamma$, filled triangles with lines - $Pn^{(E,\varepsilon)}$, filled squares with lines - $Pn^{(T,\varepsilon)}$, the left panel - solid line - $d\bar{\sigma}_P^{(E,M)}/dE_\gamma$, dotted line - $d\bar{\sigma}_P^{(T,M)}/dE_\gamma$; the right panel - solid line – $d\bar{\sigma}_P^{(E,B)}/dE_\gamma$, dotted line - $d\bar{\sigma}_P^{(T,B)}/dE_\gamma$. Initial values of width $h$ are indicated in the panels.

This figure shows that the calculations with EMPIRE and TALYS codes underestimate the values of mean gamma-ray spectra $d\bar{\sigma}_P^{(\alpha,\varepsilon)}/dE_\gamma$ in the high energy range $E_\gamma > 6.6$ MeV. For $E_\gamma < 2.0$ MeV, the theoretical calculations by the EMPIRE code are in rather close agreement with normalized mean experimental spectra. In this energy range, the calculations using TALYS code underestimate the normalized mean experimental spectra, which may be related to the use for recalculations in the low-energy range of the relationship (1). It can be seen that increasing the width $h$ leads to the reduction of the oscillations in the mean gamma-spectra $d\bar{\sigma}_P^{(\alpha,\varepsilon)}/dE_\gamma$. The calculations with the use of EMPIRE code correspond to the averaged experimental data with the averaging interval about 0.5-0.8 MeV and it seems likely that spectra by TALYS code correspond averaging with $h > 1 \div 2$ MeV.

Now the estimations of mean $\gamma$-spectra are considered as the combination of theoretical values for low gamma-ray energies and averaged experimental data in high-energy range with $E_\gamma \geq E_c = 6605.4$ keV normalized to theoretical values by formula (6), that is, combined spectra $\dfrac{d\sigma_C^{(\alpha,\varepsilon)}(E_\gamma)}{dE_\gamma}$ given by the following formula:

$$\frac{d\sigma_C^{(\alpha,\varepsilon)}}{dE_\gamma} = \frac{d\sigma^{(\alpha)}}{dE_\gamma} \cdot \theta(E_c - E_\gamma) + \frac{d\bar{\sigma}_P^{(\alpha,\varepsilon)}}{dE_\gamma} \cdot \theta(E_\gamma - E_c) \;. \qquad (8)$$

The $\theta$ is a step function (the Heaviside step function) the value of which equals to zero for negative argument and one for positive argument.

Fig. 5 compares combined spectra (8) with mean normalized spectra $d\bar{\sigma}_P^{(\alpha,\varepsilon)}/dE_\gamma$ given by (3), (7) based on different experimental data ($\varepsilon = M, B$) and calculations within EMPIRE and TALYS codes ($\alpha = E, T$).

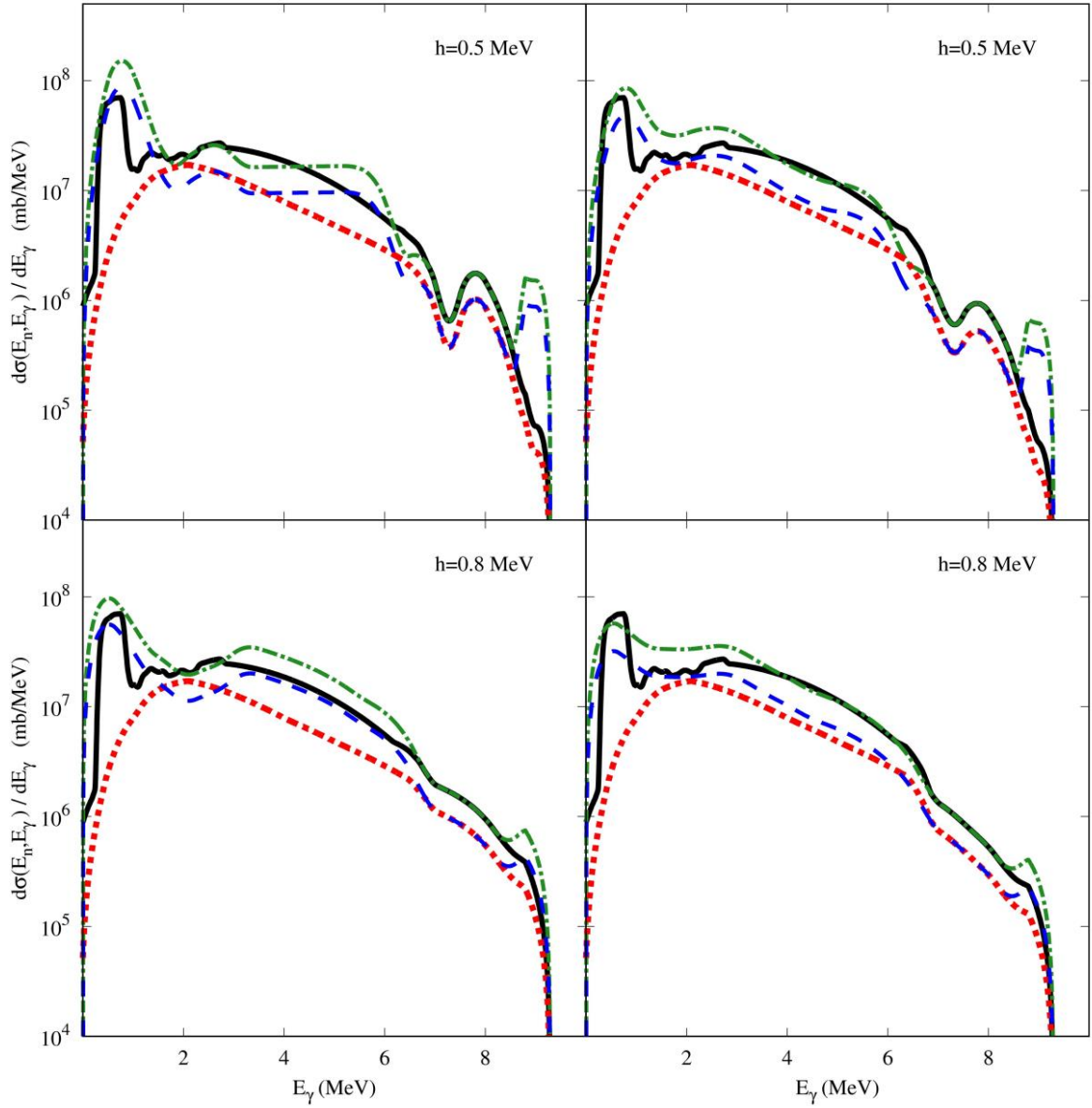

Fig. 5. Comparison of combined spectra $d\sigma_C^{(\alpha,\varepsilon)}/dE_\gamma$ with mean normalized spectra $d\bar{\sigma}_P^{(\alpha,\varepsilon)}/dE_\gamma$ for different experimental data and calculations within EMPIRE and TALYS codes ($\alpha = E, T$). The left panel shows evaluations based on the experimental data by Mheemeed et al [4] ($\varepsilon = M$), the right panel - evaluations using data Belgya et al [5] ($\varepsilon = B$). The denotations: the left panel - solid line presents $d\sigma_C^{(E,M)}/dE_\gamma$, dash-dot line - $d\bar{\sigma}_P^{(E,M)}/dE_\gamma$, dotted line - $d\sigma_C^{(T,M)}/dE_\gamma$, dashed line - $d\bar{\sigma}_P^{(T,M)}/dE_\gamma$; the right panel - solid line presents $d\sigma_C^{(E,B)}/dE_\gamma$, dash-dot line - $d\bar{\sigma}_P^{(E,B)}/dE_\gamma$, dotted line - $d\sigma_C^{(T,B)}/dE_\gamma$, dashed line - $d\bar{\sigma}_P^{(T,B)}/dE_\gamma$. Initial values of width $h$ are indicated in the panels.

The matching procedure at the construction of combined $\gamma$-spectra leads to the interface of their high energy parts with the experimental data. It is seen from Fig. 5 that combined spectra $d\sigma_C^{(\alpha,\varepsilon)}/dE_\gamma$ and mean normalized spectra $d\bar{\sigma}_P^{(\alpha,\varepsilon)}/dE_\gamma$ are in rather close agreement with the use of the data given by Mheemeed et al [4] and Belgya et al [5].

We used the normalization procedure to the two different experimental datasets from data given by Mheemeed et al [4] and Belgya et al [5]. The data from Ref.[4] were not averaged and the data from Ref.[5] were binned to 1 keV/bin.

Normalized factors were determined by equation (6) as a ratio of area values for energy-intervals with slow energy dependence of the spectra. Now we check this approach and find the normalized factors by the least-square method by fitting experimental data to theoretical calculations in the same energy interval in energy points corresponding to middle values of histogram bars used for construction of polygon, *i.e.*, by minimization of the expression

$$\chi^2 = \sum_{\substack{E_{\gamma,k} \in \Delta E_A \\ E_{\gamma,k} \in \Delta E_B}} \left( F^{(\alpha,\varepsilon)} i_k^{(\varepsilon)} / h_k - \frac{d\sigma_\gamma^{(\alpha)}(E_\gamma = E_{\gamma,k})}{dE_\gamma} \right)^2, \qquad (9)$$

where $\Delta E_A = 1879.10 \div 3238.6$ keV and $\Delta E_B = 5384.9 \div 6605.4$ keV.

New values of the factors are smaller than previous ones and for data presented for example by Mheemeed et al, they are equal to $F^{(E,M)} = 87.5$, $F^{(E,B)} = 14.9$, $F^{(T,M)} = 55.7$, $F^{(T,B)} = 10.4$. Relative uncertainties of new values in comparison with old ones are $\Delta F^{(E,M)} = (F_{new}^{(E,M)} - F^{(E,M)})/F^{(E,M)} = -12\%$; $\Delta F^{(T,M)} = (F_{new}^{(T,M)} - F^{(T,M)})/F^{(T,M)} = -8\%$.

These estimations of the normalized factor within the approach using a least square method indicate that relative uncertainty of the reconstruction procedure can be bigger than ~10% in the ranges of medium energy, where most of the unresolvable peaks are located, and at high energies. The scaling factors are also dependent on cross-section values for neutron energy 1 eV and can leads to additional uncertainty till 50%. Because we used here the values calculated by TALYS and EMPIRE codes which are approximately in 50% more than the values in other libraries (ENDF/B VIII.0, TENDL-2019 (WEB-site of NDS-IAEA); BROND-3.1 etc). This is the reason for presenting in the work the results using two sets of theoretical calculations and experimental data.

For selection of the most reliable estimation of average gamma-spectra, we calculate the mean gamma-ray multiplicities $\bar{M}_\gamma$ and the mean total capture cross sections $\bar{\sigma}_{th}$ in approximation of continuous spectra. Mean gamma-ray multiplicity $\bar{M}_\gamma$ for neutron capture at thermal energy can be calculated by the following relationship

$$\bar{M}_\gamma = U / \bar{E}_\gamma, \; U = S_n,$$

where $S_n$ is neutron separation energy ( 9.04 MeV for $^{114}Cd$ ); $\bar{E}_\gamma = \int_0^{S_n} dE_\gamma E_\gamma \cdot w(E_\gamma)$ - energy averaged over gamma-spectra with the probability $w(E_\gamma)$ of the emission of a gamma-quantum with the energy $E_\gamma$, $w(E_\gamma) = \dfrac{d\bar{\sigma}(E_\gamma)}{dE_\gamma} \bigg/ \int_0^{S_n} dE \dfrac{d\bar{\sigma}(E)}{dE}$, that is

$$\bar{M}_\gamma = \int_0^{S_n} \frac{d\bar{\sigma}}{dE_\gamma} dE_\gamma \bigg/ \bar{\sigma}_{th}, \qquad \bar{\sigma}_{th} = \int_0^{S_n} \frac{d\bar{\sigma}}{dE_\gamma} E_\gamma dE_\gamma / S_n. \qquad (10)$$

Here we omitted the indexes denoted evaluation method of gamma-spectra and took into account that the ratio of the energy weighted integral of gamma-spectrum to the separation energy gives a good approximation of the total capture cross section $\bar{\sigma}_{th}$ [10,11]. For numerical integrations, the procedure "NIntegrate" from the Wolfram Mathematica 9.0 package [12] was used with default parameters. The linear interpolation functions for spectra were taken at integration and upper limit in integrals was changed on $\tilde{S}_n = S_n + 2 \cdot h$ due to use of average spectra.

The results of the calculations are presented in the following table:

| Estimation used for gamma-ray spectrum | | $\bar{\sigma}_{th} = \int_0^{S_n} \dfrac{d\bar{\sigma}}{dE_\gamma} E_\gamma dE_\gamma / S_n$ ,b | | $\bar{M}_\gamma = \int_0^{S_n} \dfrac{d\bar{\sigma}}{dE_\gamma} dE_\gamma \bigg/ \bar{\sigma}_{th}$ | |
|---|---|---|---|---|---|
| | | $h=0.5 MeV$ | $h=0.8 MeV$ | $h=0.5 MeV$ | $h=0.8 MeV$ |
| $d\bar{\sigma}_P^{(\alpha,B)}/dE_\gamma$ | $\alpha = T$ | 24965 | 24069 | 3.98 | 3.82 |
| | $\alpha = E$ | 44659 | 43055 | 3.98 | 3.82 |
| $d\bar{\sigma}_P^{(\alpha,M)}/dE_\gamma$ | $\alpha = T$ | 27205 | 30313 | 4.33 | 3.80 |
| | $\alpha = E$ | 47099 | 52481 | 4.33 | 3.80 |
| $d\sigma_C^{(\alpha,B)}/dE_\gamma$ | $\alpha = T$ | 17911 | 18541 | 3.05 | 2.99 |
| | $\alpha = E$ | 36597 | 36720 | 3.54 | 3.53 |
| $d\sigma_C^{(\alpha,M)}/dE_\gamma$ | $\alpha = T$ | 18385 | 18541 | 3.01 | 2.99 |
| | $\alpha = E$ | 37372 | 37643 | 3.49 | 3.47 |

It should be noted that the value of the cross-section of thermal neutron capture equals $\sigma_{th} = 19964.1$ b in accordance with ENDF/B-VIII.0 library and the gamma-multiplicity calculated in [5] is equal to $\bar{M}_\gamma = 4.11$. It can be seen by comparing these quantities with table values that the most adequate values of $\bar{M}_\gamma$ and $\bar{\sigma}_{th}$ correspond to mean gamma-ray spectrum $d\bar{\sigma}_P^{(T,B)}/dE_\gamma$ which was found by averaging frequency polygon of experimental data by Belgya et al [5] ($h=0.5$ MeV) and scaling theoretical calculations within TALYS. The mean gamma-ray spectrum $d\bar{\sigma}_P^{(T,B)}/dE_\gamma$ ($h=0.5$ MeV) is preferable for use, because it leads to the better consistency of the $\bar{M}_\gamma$ and $\bar{\sigma}_{th}$ with the corresponding experimental values.

## 3. Conclusions

Mean gamma-ray spectra from $^{114}Cd$ after thermal neutron capture in $^{113}Cd$ were evaluated in units of mb/MeV. Two variants of estimation of the average gamma-ray spectrum with normalization of the experimental data were considered: I) mean spectrum for all gamma-energies was found by averaging frequency polygon for the histogram of experimental data [4,5], and II) mean spectrum was estimated as the combination of theoretical values at low gamma-ray energies and averaged experimental data in high-energy range with $E_\gamma > 6.5$ MeV.

The procedure of reconstruction of this gamma-spectra with the use of theoretical calculations and experimental data was given. This approach allows to obtain a general trend of gamma-spectrum in the middle energy range (~2.0 MeV – 6.5 MeV) and in the peak region.

Accuracy of the reconstructed high-energy part of the spectra depends mainly on the quality of experimental data as well as on theoretical approximation of the spectra in the energy ranges with smooth energy dependence within TALYS and EMPIRE codes. In these energy ranges, the calculations are practically independent of PSF and NLD expressions for codes.

The mean gamma-ray spectrum $d\bar{\sigma}_P^{(T,B)}/dE_\gamma$ ($h=0.5$ MeV) based on averaging frequency polygon for experimental data by Belgya et al [5] ($h=0.5$ MeV) and scaling theoretical calculations within TALYS is preferable for use, because it leads to the better consistency of the mean gamma-ray multiplicity and thermal capture cross-section with the corresponding experimental values.

We can not estimate the total uncertainties of theoretical calculations and experimental data. The calculations of the normalized factor within the approach using a least square method indicate that relative uncertainty of the reconstruction procedure can be bigger than ~10% in the ranges of medium energy, where most of the unresolvable peaks are located and at high energies. This is the reason for presenting in the work the results with using two sets of theoretical calculations and experimental data. The scaling factors are also dependent on cross-section values for neutron energy 1 eV and can leads to additional uncertainty till 50%. This results from using in the estimations the values calculated by TALYS and EMPIRE codes which are approximately in 50% more than the value ~ 130 b in other libraries (ENDF/B VIII.0, TENDL-2019 (WEB-site of NDS-IAEA); BROND-3.1 etc).

There are other methods for such estimations [5, 11, 13-15], for example, based on normalization using determination of the characteristics of highest energy peak with an additional implementation either unfolding procedure or relation between peak intensities and partial cross-sections. Such methods are more complicated and involve many different investigations, including the improvements in resolution and registration efficiency of the high-energy gamma-rays. It seems to us, that the proposed procedure for the evaluation of the high-energy part of thermal neutron capture gamma spectra is the simplest method for the estimation of the high-energy part of thermal neutron capture gamma spectra.